# Giant water clusters: where are they from?

T. Yakhno[1,3], M. Drozdov[2], V. Yakhno[1,3]

[1]Federal Research Center Institute of Applied Physics of the Russian Academy of Sciences (IAP RAS), Nizhny Novgorod, Russia;
[2]Institute for Physics of Microstructures RAS (IPM RAS), Nizhny Novgorod, Russia;
[3]N.I. Lobachevsky University of Nizhny Novgorod (National Research University), Nizhny Novgorod, Russia.

yakhta13@gmail.com; yakhno@appl.sci-nnov.ru

**Abstract**. A new mechanism for the formation and destruction of giant water clusters described in the literature is proposed. We have earlier suggested that the clusters are associates of liquid crystal spheres (LCS), each of which is formed around a seed particle, a microcrystal of sodium chloride. In this paper, we show that the ingress of LCS in water from the surrounding air is highly likely. When a certain threshold of the ionic strength of a solution is exceeded (for example, in the process of evaporation of a portion of water), the LCS begin to "melt", passing into free water, and the salt crystals dissolve, ensuring re-growth of larger crystals as a precipitate on the substrate. A schematic diagram of the dynamics of phase transitions in water containing LCS during evaporation is proposed.

The structure and dynamics of water has been a top debatable topic for more than a decade [1-5]. To explain a number of its anomalous properties, a two-phase water concept has been developed [6,7]. The reasons for the anomalous physical properties of water are traditionally sought at the atomic-molecular level, in the nanometer space, and on the picosecond time scale. However, recent data allow considering water as a micro-dispersed system. To the best of our knowledge, giant (millimeter-sized) clusters in a thin layer of water were detected for the first time using the IR spectroscopy [8] and were assumed to have liquid crystal nature. Subsequently, supramolecular water complexes ranging in size from 10 to 100 μm were visualized using laser interferometry [9,10] and small-angle light scattering [11,12]. The nature of these clusters is still controversial. The authors of [13] used dielectrometry and the resonance method to show that with an increase in the frequency of reactive current from 1 to 300 kHz, the electrical capacitance of distilled water decreases several fold. With increasing concentration of NaCl aqueous solutions, their electrical capacity increases several fold. The authors suggested that these changes in distilled water are due to the presence of interconnected associates in it. Changes in NaCl solutions, according to the authors, depend on the ratio of the number and size of water associates and the degree of hydration of ions. A decrease in the content of water aggregates ($r \leq 5$ μm) under the action of an external constant magnetic field with a magnetic induction of 1.5 T for 30 minutes was also noted in [12].

In our previous studies by the method of acoustic impedancemetry, we identified slow (near-hour) self-oscillation processes in colloidal liquids [14] and proposed a mechanism for their implementation [15]. It is based on periodic phase transitions between the free and bound (liquid crystal) water of the hydration shells of the dispersed phase, which manifests itself in fluctuations in the density of the liquid. These transitions are regulated and coordinated in the whole volume by osmotic pressure changing as a result of these transitions. Morphologically, these processes manifest themselves as the growth and destruction of spherical structures of micron size (50 - 250 μm), transparent in a liquid medium and visible in it due to the "contouring" by the particles of the dispersed phase. In-phase with the growth and destruction of water microstructures, the surface tension of the test solution fluctuates at the "liquid – air" boundary. Liquid crystal water spheres exhibit the properties of a viscous liquid and evaporate at temperatures above 200 ° C [16].

The fact of formation of a quasi-crystalline phase of water at hydrophilic surfaces (both moving [17-19] and stationary [20-22]) has been repeatedly recorded over the past 100 years and experimentally confirmed by many independent researchers. Those discoveries have made an immeasurable contribution to the development of hydrodynamics and biology. To study the phenomenon in laboratory, Nafion - fluorocarbon polymer membranes, including hydrophilic sulfone groups, are commonly used as a hydrophilic surface [22,23]. When immersed in water, an "exclusion zone" (EZ) – a water layer with a more dense molecular packing than free water – is formed in 10 to 60 minutes. The width of this layer can reach 500 μm. At the same time, other ions and dispersed particles are displaced from the close-packed water zone. The growth of "exclusion zones" near the surface of metals and glass was also noted [23,24]. In recent years the concept of the molecular structure of the "exclusion zone" has evolved from multi-layered packing of water dipoles [21,22] to a layered structure of flat sheets of hexagonal cells formed by water molecules, where the sheets, unlike crystalline ice, are interconnected by weak electrostatic interaction rather than by rigid hydrogen contacts [23,25]. This ensures that the sheets glide relative to each other, as a result of which the "exclusion zone" behaves like a viscous liquid. It is noteworthy that, in accordance with the study [26], during the melting of ice, an LC phase of water first forms and then free water. Based on the results of our work [14–16], we continued to study the origin and areas of existence of LCS when considering the phase transformations of water in the process of its evaporation. The structure and atomic composition of materials were analyzed by the method of X-ray diffraction and secondary ion mass spectrometry (SIMS).

1. Materials and methods

Tap, distilled and deionized water and aqueous solutions were used in the study. The experiments were performed under laboratory conditions at T = 22°-24° C, H = 73% -75%. The electrical conductivity of the solutions was measured with a MARK-603 conductometer (Russia). "Crushed drop" preparations and smears for microscopic observations were prepared as previously described [15], using "ApexLab" slides (Cat. No. 7105) and "ApexLab" cover slides (24x24 mm). The samples were examined under a Levenhuk microscope with a computer-coupled video camera Levenhuk C-1400 NG using the ToupView program.

Air was passed through the water using a uniStar AIR 1000-1 aquarium compressor (2.5 W, 72 l / h) for 10 min. The X-ray diffraction experiment was performed on a Bruker D8 Discover X-ray diffractometer. The survey was carried out in a sliding incidence geometry (angle of incidence - 3°) with a Gebel mirror and a 0.6 mm gap on the primary beam. A 2Θ scan was recorded with a Soller gap in front of the detector.

SIMS measurements were made on a TOF.SIMS-5 installation with a time-of-flight mass analyzer. We used cluster probe ions $Bi^{3+}$ with an energy of 25 keV and a current of 1 pA in a single pulse. Separately, negative and positive secondary ions were recorded. Mass spectrometry of secondary ions (ionized sputtering products) allows surface and volumetric analysis of element concentrations to be performed. Using SIMS, a qualitative analysis of microstructures was carried out with reliable identification of all elements present with a sensitivity level of $N\alpha > 10^{14}-10^{16}$ at / $cm^3$. In addition, the image of the surface in the secondary ions (lateral resolution of 0.1-0.50 microns) was obtained. The measurements were performed in the static SIMS mode, which ensures the nondestructive character of surface analysis.

2. Results and Discussion

In our earlier work we showed based on optical microscopy that under natural conditions water is a dispersed system, in which the dispersed phase is represented by salt microcrystals surrounded by thick hydrated shells – liquid crystal spheres [27]. LCS were also present on the surface of dry glass and plastic. However, the origin of salt microcrystals in a liquid medium remained unknown. After free evaporation of a layer of distilled water ~ 2 mm thick from a glass Petri dish (d = 9 cm) under room conditions, a loose sediment remained at the bottom, which eventually turned into an array of salt crystals (Fig. 1). The electrical conductivity of distilled water used in this experiment was 36.5 µS / cm.

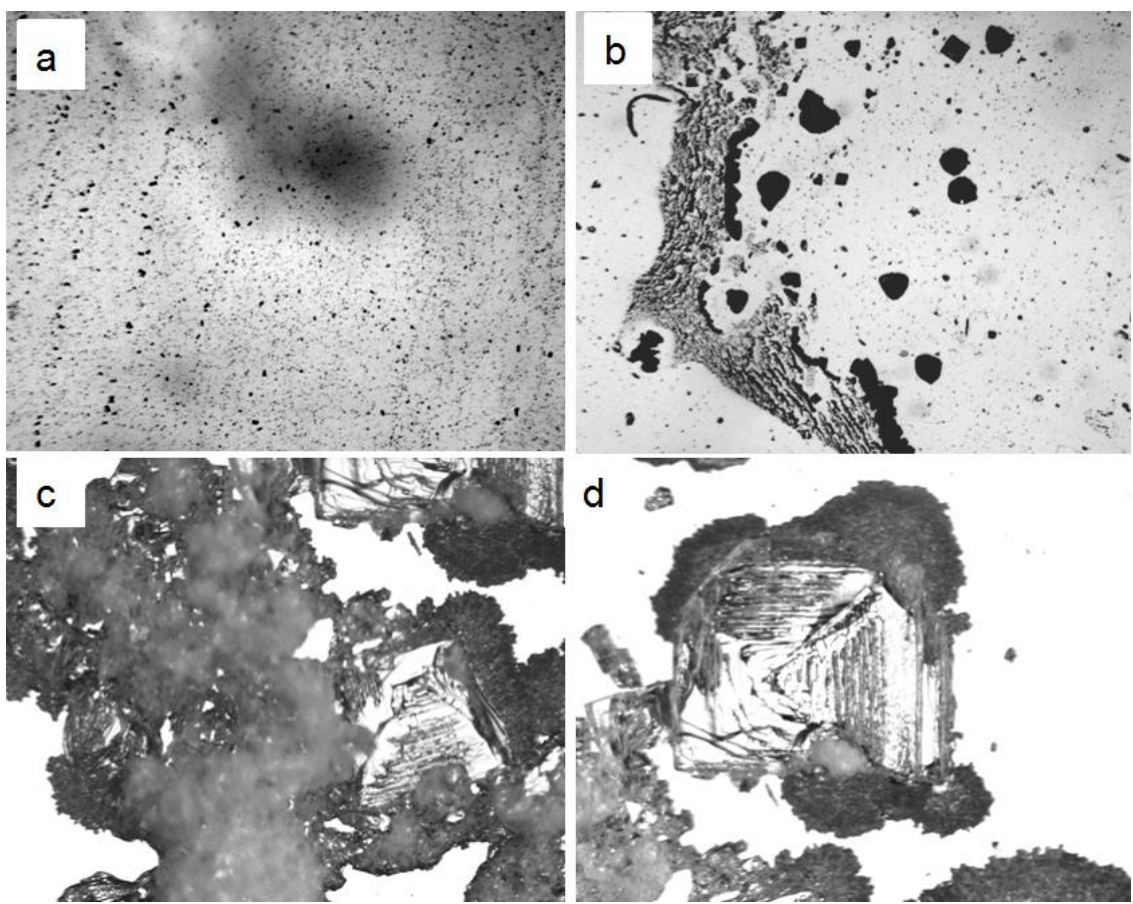

**Fig. 1.** Structures formed after drying of distilled water in a Petri dish (layer thickness ~ 2 mm): a - 7 days after the start of the experiment, b - 14 days after; c, d - large crystals grown from the sediment mass (a) one week after scraping it with a scalpel into a single mass. The width of the frames a, b - 3 mm, c, d - 1 mm.

Analysis of the crystals using an X-ray diffractometer unambiguously confirmed that it is sodium chloride (Fig. 2). That is, NaCl microcrystals were really "priming" for the formation of LCS.

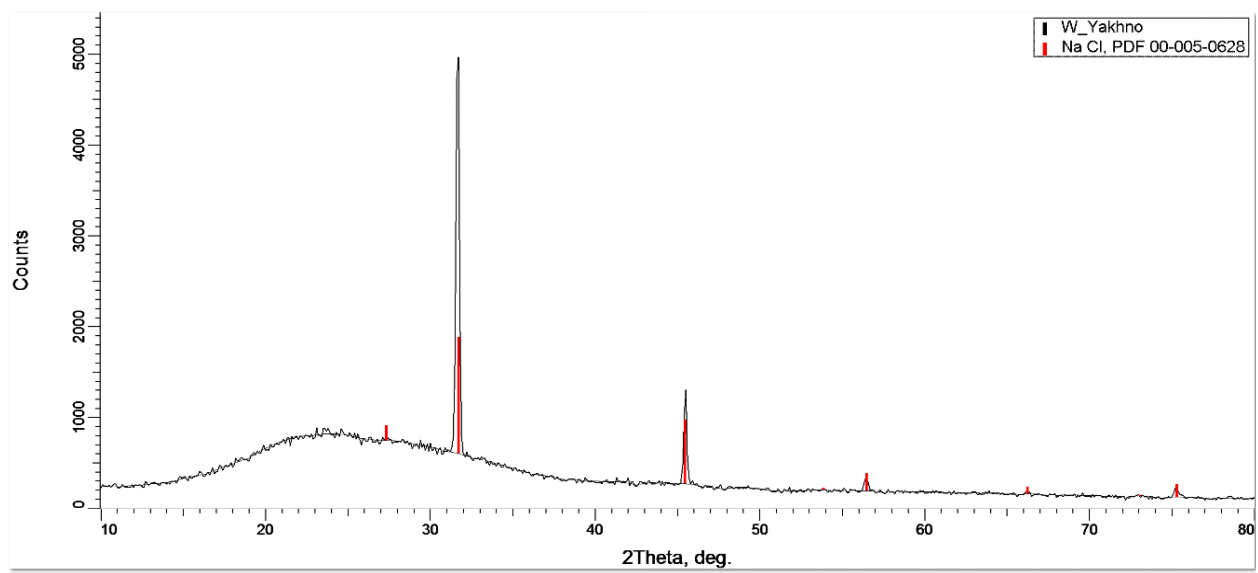

**Fig. 2.** Diffraction pattern of crystals formed after evaporation of distilled water.

To test the possibility of salt ingress from the ambient air into water, the following experiment was carried out. 50 ml of distilled water from the same container was poured into two identical clean glass beakers. The end of a plastic tube connected to an aquarium compressor was immersed into one of the beakers. The input end of the tube was placed above the laboratory table in the working area. Laboratory air was passed through the water for 10 min, at a rate of 72 l / h (before the experiment, the compressor was idling for 20 min to clean air paths from possible internal contaminants). Thus, 12 l of air passed through 50 ml of water during the experiment. Both beakers (control and experimental) were left on the table at room conditions, covered with a flat lid, for a week to restore the structural balance. A week later the electrical conductivity of the control sample was 4.1 µS/cm, and of the experimental one 5.4 µS/cm. The microscopy of the droplets of the water samples tested on the glass slide also revealed the difference in the content and structure of the sediment (Fig. 3).

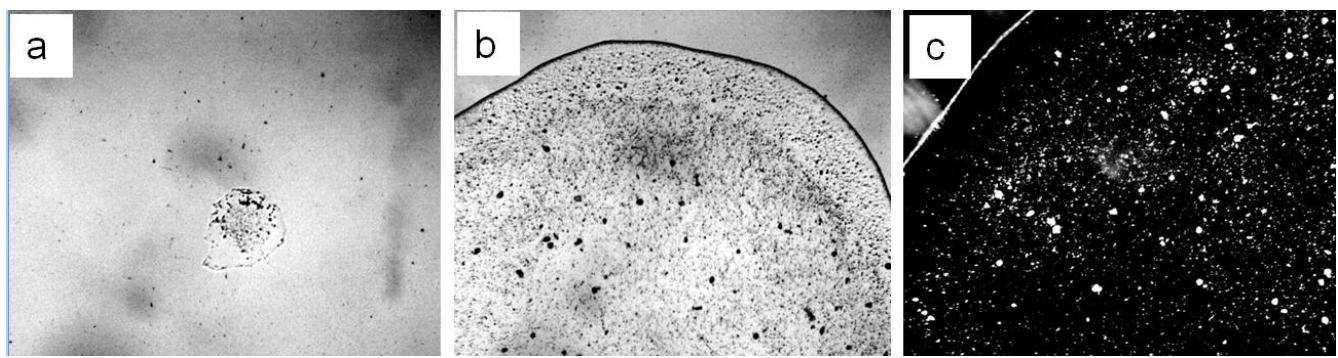

**Fig. 3.** Microphoto of water droplets dried on glass: a - control; b, c - after passing the air (c - dark-field image). Frame width: a, b - 3 mm, c - 1 mm.

Obviously, the passage of air has led to severe water pollution. Moreover, salts constituted a large percentage of the microimpurities, which ensured an increase in the electrical conductivity of the sample. Thus, the aerosol origin of LCS is quite likely.

Secondary ion mass spectrometry was used to measure static mass spectra from the surface of a clean silicon wafer before and after evaporation on it of a drop of deionized water stored in the laboratory. In the dried drop of water, a strong increase in the intensity of all lines containing carbon and sulfur was noted (Fig. 4, 5).

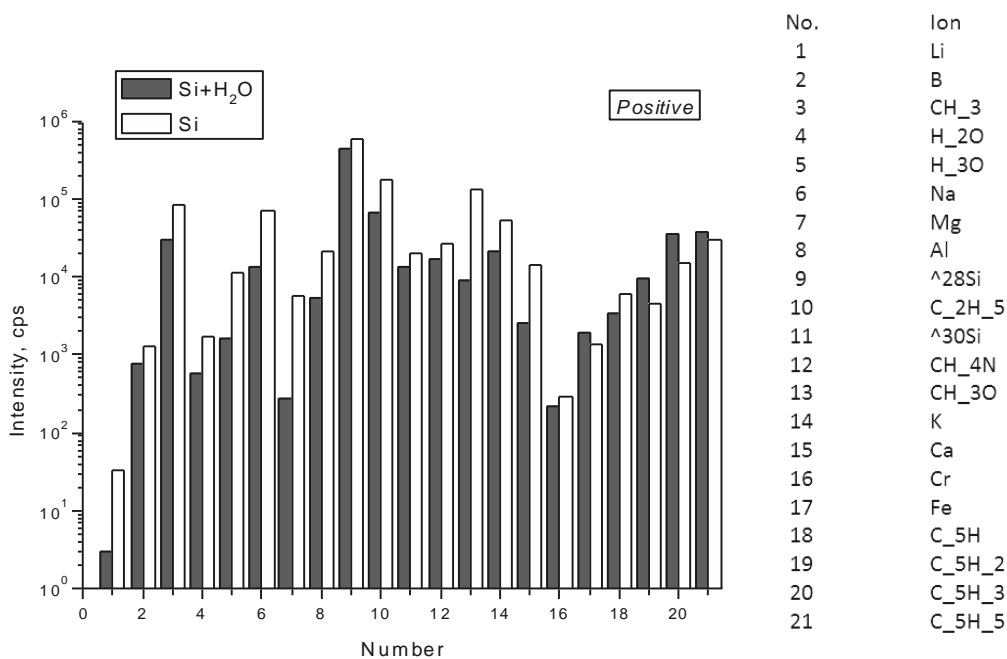

**Fig. 4.** Intensity of the lines of positive secondary ions on the surface of silicon wafer before and after evaporation of a drop of deionized water on it (light and dark bars, respectively).

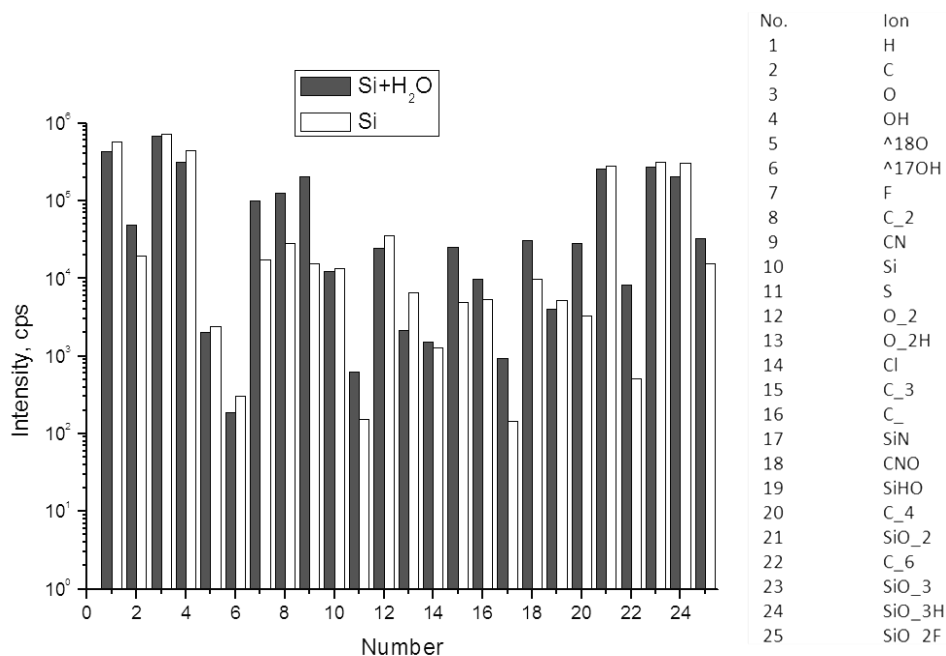

**Fig. 5.** Intensity of the lines of negative secondary ions on the surface of silicon wafer before and after evaporation of a drop of deionized water on it (light and dark bars, respectively).

The radial distribution of ions over the spot of the dried drop was uneven (Fig. 6). In the central zone of the spot of the dried drop, the content of carbon, sulfur and chlorine prevailed, whereas lithium, sodium, potassium, magnesium, calcium and chlorine ions were more concentrated at the periphery. The presence of carbon and sulfur in a dried drop may indicate the presence of soot, which, like salt, could get into the water from the air.

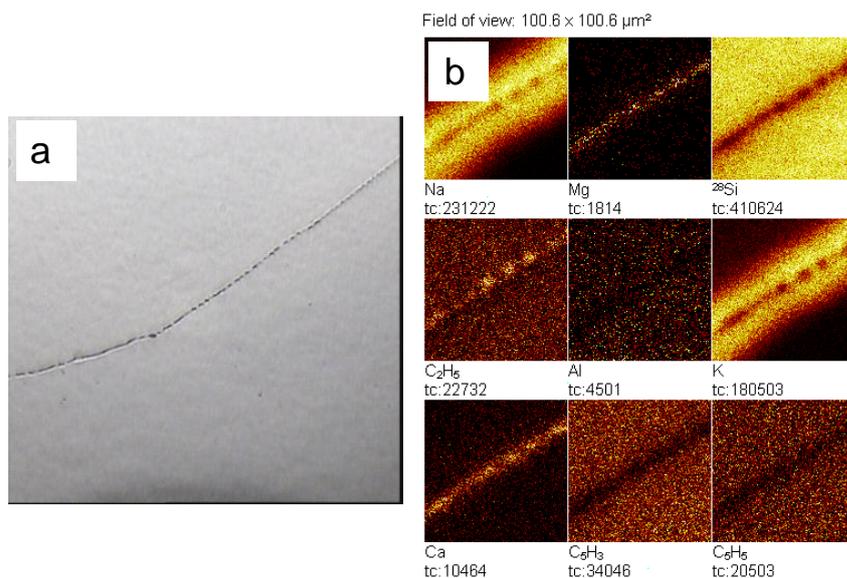

**Fig. 6.** The surface of the dried drop according to SIMS: a - image of a portion of the surface of Si on which there is a fragment of the perimeter of the dried drop of water (obtained by the integrated video camera of the TOF.SIMS-5 installation; frame size 1000x1000 μm); b - lateral distribution of positive secondary ions on the Si surface, including the perimeter of the dried drop.

The dilution effect of small amounts of salts on colloidal structures has been actively discussed since the middle of the last century ([28], p. 50). Let us consider in more detail the evolution of structures on a glass slide when a portion of tap water (1 ml) containing micro-impurities dries out (Fig. 7). At the beginning of the drying of the water sample, part of the LCS, as a colloidal phase, is transferred by a capillary flow to the three-phase boundary [29] (Fig. 7, a, b). A progressive decrease in the concentration of free water is accompanied by an increase in the ionic strength of the solution. At the same time, part of the liquid crystal water surrounding the salt "seed" under the action of osmotic pressure melts, turning into free water and creating conditions for the growth of crystals with further evaporation of free water.

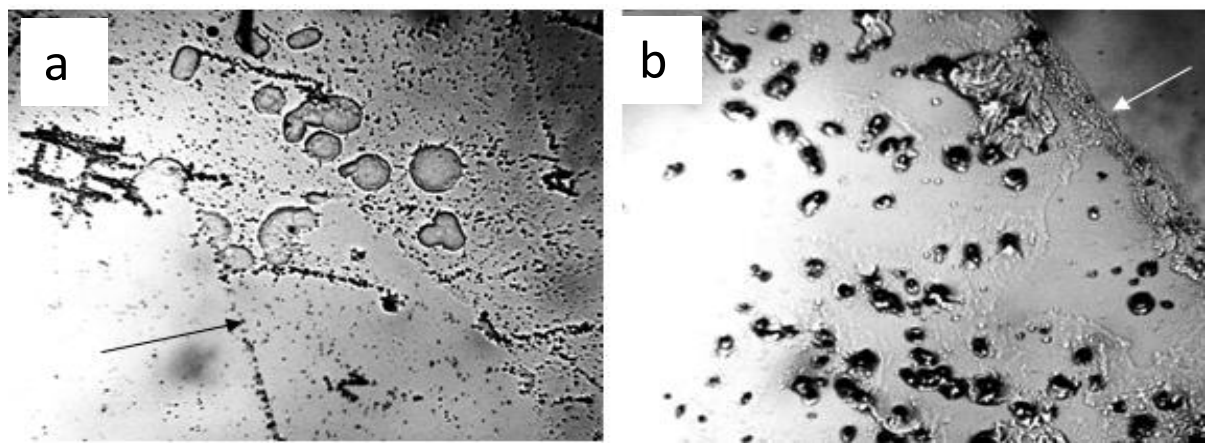

**Fig. 7.** Microscopic picture of structures in tap water dried on glass (1 ml) 5 days after application: a - melting of LCS and merging of the formed liquid droplets; b - growth of larger salt crystals. Frame width - 1 mm.

After drying of 1 ml of distilled water with an electrical conductivity of 36.5 μS / cm, the beginning of salt erosion of the LCS at the interface was observed on a glass slide (Fig. 8, a, b). In the dried sample of water with a lower electrical conductivity (14.0 μS / cm), large aggregates of LCS settling on the substrate were found (Fig. 8, c, d).

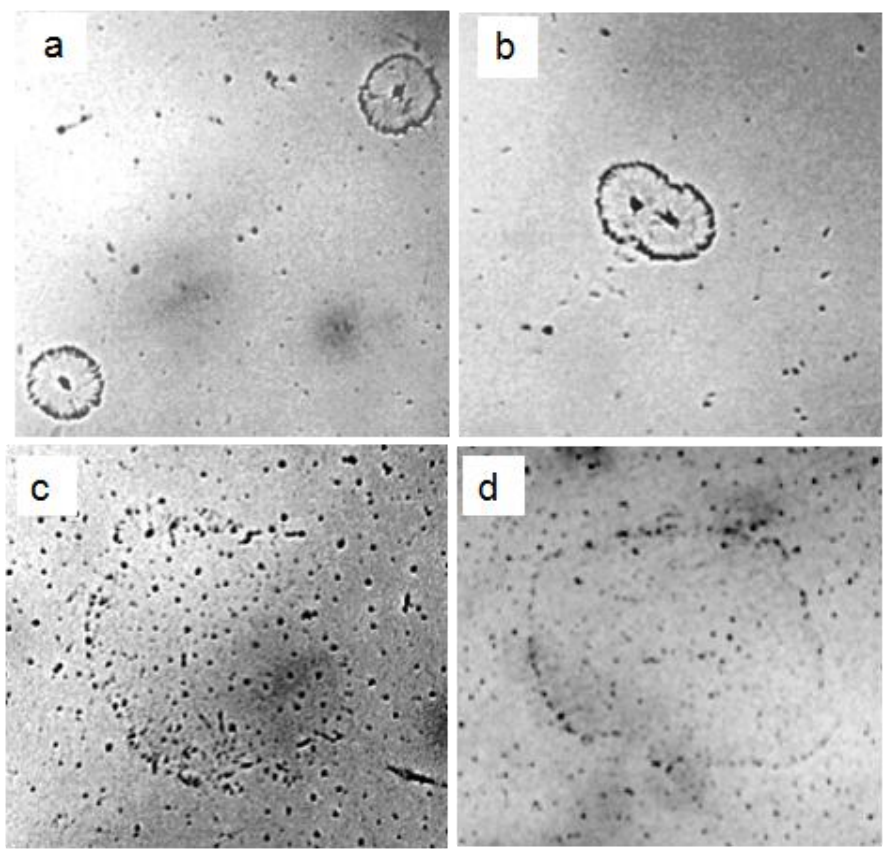

**Fig. 8.** Structures on the glass after drying 1 ml of distilled water with electrical conductivity of 36.5 μS/cm (a, b) and 14 μS/cm (s, d). a, b - onset of salt erosion of LCS, which is indicated by the radial striation and scalloped edges of the structures; c, d - aggregates of FSW deposited on glass from the dried layer of liquid water. The width of each frame is 1.0 mm.

Experimental data and available literature sources presented in this work allowed us to present a schematic diagram of the dynamics of phase transitions in water containing FSL with NaCl microcrystals as a "seed" when it evaporates from a solid substrate (Fig. 9).

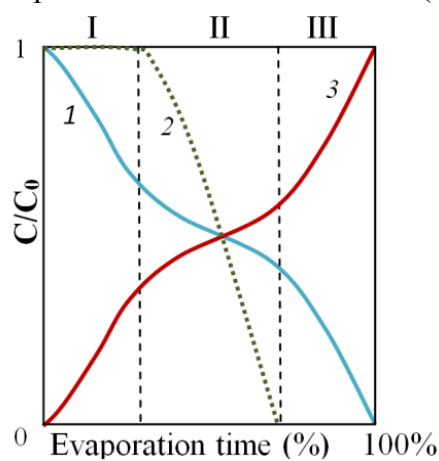

**Fig. 9.** Schematic diagram of the dynamics of phase transitions in water containing LCS with NaCl microcrystals as a "seed" during its evaporation from a solid substrate, in the "Relative concentration - Time (%)" coordinates. Stage I - evaporation of free water and increase in osmotic pressure; Stage II - phase transition of LC-water into free water and reducing relative rate of evaporation; Stage III - growth of NaCl crystals. 1 - relative concentration of free water; 2 - relative concentration of LC-water; 3 - relative salt concentration.

3. Conclusion

The results of the study and analysis of the literature enable us to make the following conclusion. Aerosol contamination of water, which is almost impossible to avoid in real life, is accompanied by the appearance in the liquid medium of "giant clusters" - micro aggregates of LCS, each of which is formed around a salt microcrystal. Salt contained in LCS does not dissolve, as LC-water has insufficient dissolving ability. At a certain stage of water drying, when the osmotic pressure reaches a critical level, the LCS are eroded and "melt". The released microcrystals of salt obtain an opportunity for further growth and "Ostwald ripening". The methodological approach we used allows us to consistently explain the mechanism of the formation of "giant clusters" in bulk water described by other researchers. It is known that non-reagent energy impact on water leads to a uniform reversible change in its physicochemical properties: an increase in pH, a decrease in surface tension, an increase in the speed of ultrasound. All these changes can be associated with the destruction of the LCS aggregates (an increase in the dispersion and the interphase surface adsorbing protons) and a slow return to the initial state after the termination of the energy source. This assumption will be the subject for further research. The coexistence of liquid crystal and free phases of water in dynamic equilibrium explains, in our opinion, the anomalous characteristics of water as a physical object.

The work was supported by the Ministry of Education and Science of Russia (Project No. 14.Y26.31.0022). The work was also carried out within the framework of the state assignment of the IAP RAS (project No. 0035-2014-0008). The studies were carried out using the equipment of the Center for Collective Use "Physics and Technology of Micro- and Nanostructures of the IPM RAS. The authors are deeply grateful to P.A. Yunin for conducting the crystallographic research and discussion, A.G. Sanin - for technical assistance in organizing the experiment.